\documentstyle[12pt]{article}
\newcommand{\sign}{\mathop{\rm sign}\nolimits}
\newcommand{\const}{\mathop{\rm const}\nolimits}

\newcommand{\textlineskip}{\baselineskip=13pt}

\title{Global Properties of Exact Solutions in Integrable
Dilaton-Gravity Models}
\author{A.T. Filippov, V.G. Ivanov\\
{\it Joint Institute for Nuclear Research, Dubna, Russia} \\
{\small e-mail: filippov@thsun1.jinr.ru, vivanov@thsun1.jinr.ru}}
\date{}  

\def\metl{g_{ij}}

\def\drho{\dot{\rho}}
\def\dphi{\dot{\phi}}
\def\dpsi{\dot{\psi}}

\def\lf{(\log{f})_{uv}}

\def\be{\begin{equation}}
\def\ee{\end{equation}}
\def\bea{\begin{eqnarray}}
\def\eea{\end{eqnarray}}

\begin{document}

\maketitle

\begin{center}
\bf{Abstract}
\end{center}
{\small
\noindent
Global canonical transformations to free chiral fields are
constructed for DG models minimally coupled to scalar fields.
The boundary terms for such canonical transformations
are shown to vanish in asymptotically static coordinates
if there is no scalar field.}

\vspace*{1pt}\textlineskip
\section{Introduction}
\vspace*{-0.5pt}
\noindent
Here we consider a special class of 1+1 dimensional 
dilaton - gravity - matter models (DGM)~\cite{Last}.
Gravitational variables in these models are the metric tensor $g_{ij}$ and
the scalar dilaton field $\phi$. The matter is represented by a scalar field $\psi$,
which is minimally coupled to the gravitational variables~\cite{CGHS}.
The Lagrangian of these models may be written in the form
\be
{\cal L} = \sqrt{-g} \, [\phi R(g) + V(\phi) - g^{ij} \psi_i \psi_j/2] \, ,
\label{WI}
\ee
where $R$ is the scalar curvature, the subscripts denote
partial derivatives ($\phi_i = \partial_i \phi$, etc.),
except when used in $\metl$.
Many models (describing black holes, strings, cosmologies, etc.)
can be written in the form (1).
The formal exact solutions for some classes
of these models can be obtained by B\"acklund transformations to
free fields~\cite{Last}.
Recipes to find canonical transformations to free field
for the exactly integrable models were suggested in~\cite{Cruz}.

In what follows we study  the model in the
conformal flat, light - like metric
$ds^2 = -4f(u,v)du dv$, in which $R =(\log{f})_{uv}/f$.
Then the general covariance is lost but there exists residual
covariance under reparametrization of light - cone coordinates
$u=a(\bar{u}),\;\, v=b(\bar{v})$,
where $a(\bar{u})$ and $b(\bar{v})$ are arbitrary {\it monotonic}
coordinate functions.

\vspace*{1pt}\textlineskip
\section{Equations of Motion}
\vspace*{-0.5pt}
\noindent
The equations of motion can be derived by varying ${\cal L}$ with respect
to the variables $\metl$, $\phi$, and $\psi$,
\be
f (\phi_i /f)_i = - (\psi_i)^2/2 \; , \;\;(i = u,v) \, ,
\label{5}
\ee
\be
\phi_{uv} + fV(\phi) = 0, \;\;
\label{6}
\ee
\be
\lf + fV'(\phi) = 0 \, ,
\label{7}
\ee
\be
\psi_{uv} = 0 \, .
\label{8}
\ee
These equations are not independent.
Thus, eqs.(\ref{5})-(\ref{7}) imply eq.(\ref{8}).
A useful corollary is the following:
we only need to solve eqs.(\ref{6}) and (\ref{7}),
then the scalar field can be derived from eqs.(\ref{5})
while eq.(\ref{8}) is automatically satisfied.

If the scalar field is constant, we can find the general solution to
eqs.(\ref{5})-(\ref{7}) in terms of one free field~\cite{Last}.
Indeed, from  the constraints~(\ref{5}) it follows that
$$
f/\phi_u=b'(v)\, , \;\; f/\phi_v=a'(u)\, ,
$$
where $a(u)$ and $b(v)$ are arbitrary (smooth) functions. Thus
\be
\phi=F(\tau)\, , \;\; f=F'(\tau) a'(u) b'(v)\, , \;\;\;
\tau \equiv a(u)+b(v)\, ,
\label{f18}
\ee
where $F(\tau)$ is an arbitrary function of the free field $\tau(u,v)$.
This means that the general dilaton gravity (coupled to Abelian
gauge fields) is a topological theory and can be dimensionally reduced
to a $1+0$ dimensional theory~\cite{Banks},~\cite{Last},~\cite{VDA},~\cite{Strobl}.

To obtain the exact form of the function $F(\tau)$,
we define a new function $N(\phi)$ by $N'(\phi)=V(\phi)$.
The function
$M \equiv F'(\tau) + N(\phi) = \phi_u \phi_v / f(u,v) + N(\phi)$,
is locally conserved, i.e. $M_u=0$ and $M_v=0$.
Thus, we find
\be
\tau=\int [M-N(\phi)]^{-1}\, d\phi\, ,
\ee
and this implicitly defines the general  solution in terms of the free field
$\tau=a(u)+b(v)$.
If $N(\phi_0)=M$, $f(a,b)=F'(\tau)=M-N(\phi_0)=0$,
the function $\tau(\phi)$ is infinite at $\phi=\phi_0$,
and there is a horizon at $\tau=\infty$.

If the functions $a(u)$ and $b(v)$ are monotonic, the transformation
from variables $(u, v)$ to variables $(a , b)$ can be used.
The solution is static in the coordinates $(a,b)$
(it depends on $\tau = a+b$ only).

The general DGM~(\ref{WI}) is not integrable. However, there exists a class 
of the potentials $V(\phi)$ for which the general solution to
eqs.(\ref{5})-(\ref{7}) can be found in terms of two free fields, 
or equivalently, in terms of two pairs of chiral functions.
In what follows
they are denoted by $a(u)$, $b(v)$, $A(u)$, and $B(v)$.

\vspace*{1pt}\textlineskip
\section{Canonical transformations to free fields}
\vspace*{-0.5pt}
\noindent
To study quantum properties of the models, it is useful
to find a canonical transformation to free fields.
With this aim, we rewrite the model in the canonical form.
Let us write the two-dimensional metric in the form
\be
ds^2=e^{2\rho} \left[ -\alpha^2\, dt^2+ (\beta\, dt+ dr)^2 \right]\, ,
\label{metr}
\ee
where $\alpha(u,v)$ and $\beta(u,v)$ play the role of the lapse
function and of the shift vector, respectively; $\rho=(\log f)/2$.
We denote the derivative with respect to $t=u+v$ by the dot,
and the derivative with respect to $r=u-v$ by the prime,
\be
\partial_t = (\partial_u+\partial_v)/2 \, ,\;\;\;
\partial_r = (\partial_u-\partial_v)/2 \, .
\ee
When, using~(\ref{metr}), the action~(\ref{WI}) can be
written in the Hamiltonian form (see e.g.~\cite{Jackiw})
\be
S=\int d^2x \left( \pi_\rho \drho + \pi_\phi \dphi + \pi_\psi \dpsi
-\alpha H- \beta P\right),
\ee
where $\pi_\rho=-2\dphi$, $\pi_\phi=-2\drho$, and
$\pi_\psi=\dpsi$
are the conjugate momenta to $\rho$, $\phi$, and $\psi$;
$\alpha$ and $\beta$ are the Lagrangian multipliers;
$H$ and $P$ are the constraints:
$$
H=- \pi_\rho \pi_\phi /2 +2(\phi''-\phi' \rho') - e^{2\rho} V(\phi)
+(\pi_\phi^2 + \psi'^2)/2 \, ,
$$
$$
P=\rho'\pi_\rho-\pi'_\rho + \phi'\pi_\phi+ \pi_\psi \psi' \, .
$$

To get a canonical transformation to free fiels let us
rewrite the symplectic 2-form
\be
\omega=\int dr \, (\delta\rho \wedge \delta\pi_\rho +
\delta\phi \wedge \delta\pi_\phi + \delta\psi \wedge \delta\pi_\psi)
\label{omega}
\ee
in the new parametrization given by
the exact solution of the Lagrange equations.
In the exact solution, written in terms of free fields,
we will substitute the free chiral functions
by arbitrary functions and the derivatives with respect to $u$, $v$
by $\partial_r$
$$
\begin{array}{ll}
a(u) \to a(u,v)\, ,\;\;\; & b(v) \to b(u,v)\, , \\
A(u) \to A(u,v)\, ,\;\;\; & B(v) \to B(u,v)\, , \\
\partial_u \to 2\partial_r\, ,\;\;\;
& \partial_v \to -2 \partial_r\, ,
\end{array}
$$
since $2\partial_r=\partial_u-\partial_v$
(This recipe was proposed in~\cite{Cruz}.).

For the CGHS model~\cite{CGHS}, $V=g=\const$, the global solution 
can be rewritten as
\be
\begin{array}{cclccl}
\rho&=&\log (4 a' b')/2\, , \;\;
& \phi&=& g a b + A + B \, ,\\
\pi_{\rho}&=& -2g (a'b-ab')-2A'+2B'\, ,\;\;
& \pi_{\phi}&=& -\log (a')' + \log (b')' \, .
\label{podstanovka}
\end{array}
\ee
Thus, we get
\be
\begin{array}{rcl}
\omega &=& 2\int dr \left[ -\delta (\log a') \wedge \delta A' +
\delta(\log b') \wedge \delta B' \right] + \omega_{b_{0}}   \\
&=& 2\int dr \left[ \delta(\log (A'/a')) \wedge \delta A' -
\delta(\log (B'/b')) \wedge \delta B' \right] + \omega_{b_{0}} \, ,
\end{array}
\ee
Here $\omega_{b_{0}}$ is the boundary term,
which arises from integration by parts,
\be
\omega_{b_{0}}=\int d \left[ - \delta\phi \wedge \delta \log (a'/b')
-2g \delta a \wedge \delta b \right] \, ,
\label{obCGHS}
\ee
and $\phi$ is defined in~(\ref{podstanovka}).
To rewrite the constraints $C^{\pm}=\pm(H \pm P)/2$
in the simple form~\cite{Jackiw},~\cite{Kuchar},
\be
C^{\pm}=\pm X^{\pm'} \Pi^{\pm} + (\pi_{\psi} \pm \psi')^2/4 \, ,
\label{standard}
\ee
we can choose the canonical coordinate $X^{+}$ and
momentum $\Pi^{+}$ in one of the following four forms ($i=1,...,4$):
\be
\begin{array}{rccccc}
X^{+}_{i}&=& a\, , \;\;\; &(A'/a')\, , \;\;\; &
A\, , \;\;\; &\log(A'/a') \, ,\\
\Pi^{+}_{i}&=& 2(A'/a')'\, ,\;\;\; & 2a' \, ,\;\;\; &
2(\log (A'/a'))'\, ,\;\;\; &2A'\, .
\end{array}
\ee
The canonical coordinate $X^{-}$ and momentum
$\Pi^{-}$ can be chosen independently of $X^{+}, \; \Pi^{+}$.
Altogether this gives $4 \times 4 = 16$ different canonical
transformations to the chiral coordinates and momenta
\be
\omega_{ij}=\int dr \, (\delta X^{+}_{i} \wedge \delta \Pi^{+}_{i} +
\delta X^{-}_{j} \wedge \delta \Pi^{-}_{j} +
\delta\psi \wedge \delta\pi_\psi) +
\omega_{b_{0}} + \omega^{+}_{bi} + \omega^{-}_{bj} \, ,
\ee
where
$\omega_{b_{0}}$ is given by~(\ref{obCGHS}), $i,j=1,\ldots,4$
\be
\begin{array}{rclrcl}
\omega^{+}_{b1}&=&
-2 \int d \left[ \delta a \wedge \delta (A'/a') \right] \, , \;\;\; &
\omega^{+}_{b2}&=&0\, ,\\
\omega^{+}_{b3}&=&
-2 \int d \left[ \delta A \wedge \delta \log(A'/a') \right] \, , \;\;\; &
\omega^{+}_{b4}&=&0\, ,
\end{array}
\ee
and $\omega^{-}_{bi}$ are obtained from $\omega^{+}_{bi}$
by substituting $a$ with $b$ and $A$ with $B$.

In what follows we choose
$i=j=1$, $X^{+}=a$, $X^{-}=b$.
This choice of canonical coordinates seems natural because
$a$ and $b$ are the global coordinate functions on
the maximally extended solution.
Thus, the canonical transformation is global.
Below we will write the global canonical transformations
for other DG models.

Since the functions $X^{\pm}$ are monotonic coordinate functions,
we can include
$X^{\pm'}$ into the Lagrange multipliers~\cite{Kuchar},~\cite{VDA_last},
and the new constraints will be
\be
{\bar C}^{\pm} \equiv C^{\pm}/X^{\pm'}
=\pm \Pi^{\pm} + (\pi_{\psi} \pm \psi')^2/(4X^{\pm'}) \, .
\label{newc}
\ee

We can simplify the form of the boundary term by
choosing a special asymptotic behavior of the coordinate functions
$a$ and $b$.
If we choose, up to arbitrary linear transformations,
the static form of the coordinate functions,
$$
a= \pm e^{\pm u}\, , \;\;\;
b= \mp e^{\mp v}\, ,
$$
on the left and the right space-like infinities, respectively,
and require that there are constant scalar fields at infinities
($A' \to 0 \, ,\;\; B' \to 0 \;\; \mbox{for} \;\; r \to \pm \infty$),
we will completely get rid of the boundary term.
Moreover, in this way we can get rid of the boundary term for
any DG model with constant scalar fields.
Indeed, for any model ofthis kind
it is easy to prove the following statement.
For all the solutions in the static parametrization $(a,b)$ 
the boundary term is
\be
\omega_{b}=-\int d \left[
\delta \phi \wedge \delta \log(a'/b')
\right] \, .
\ee

\section{Conclusions}
Starting from global solutions of exactly integrable models
we have constructed global canonical transformations to free chiral fields.
Monotonic chiral coordinate func\-tions of the global solution
(in the conformal metric) can be chosen as new canonical coordinates.
This choice of canonical coordinates
guarantees the global character of the canonical transformation.
We can rewrite the constraints as linear functions
of momenta canonically conjugate to coordinate functions.
To get rid of the boundary term in the canonical transformation
we demand the static form of the solution (with a constant scalar field)
at spatial infinities.

Let us compare the constructions of the canonical transformations
to free chiral fields for different DG models.
The CGHS model is the simplest integrable model.
It has only one static solution.
The coordinate functions of the global solution were chosen as
canonical coordinates. We get rid of the boundary term of
the canonical transformation,
if the global coordinate functions are chosen static at spatial infinities,
and the scalar field is constant.

The generalized CGHS model ($V=g_1+g_2 \phi$)
has several static solutions.
The global static solution can be chosen as a vacuum and
may be used in the boundary conditions to get rid of the boundary term.
To study the solutions obtained
from the periodic static solutions with horizons,
we have to restrict the manifold to a cylinder.
Then we may 
completely get rid of the boundary term
by choosing periodic boundary conditions for free chiral fields~\cite{Cruz}.

For the bi-Liouville model ($V=(g_{+}e^{g\phi}-g{-}e^{g\phi})/g$)
minimally coupled to one scalar field
we can construct canonical trans\-forma\-tions to free chiral fields
and choose the boundary conditions similarly to the CGHS
(for $\sign\{g_{+}g_{-}\}=-1$) or the generalized CGHS
(for $\sign\{g_{+}g_{-}\}=1$) and get rid of the boundary term.

\vspace*{20pt}
\noindent
This investigation was partially supported
by the Russian Foundation for Basic Research (project 97-01-01041),
and by INTAS (project 93-127-ext).

\end{document}